\documentclass[sigconf]{acmart}

\copyrightyear{2018} 
\acmYear{2018} 
\setcopyright{rightsretained} 
\acmConference[WebSci '18]{10th ACM Conference on Web Science}{May 27--30, 2018}{Amsterdam, Netherlands}
\acmBooktitle{WebSci '18: 10th ACM Conference on Web Science, May 27--30, 2018, Amsterdam,
Netherlands}\acmDOI{10.1145/3201064.3201097}
\acmISBN{978-1-4503-5563-6/18/05}

\usepackage[english]{babel}

\usepackage[T2A,T1]{fontenc}

\usepackage{booktabs}

\usepackage{graphicx}
\usepackage{subcaption}

\usepackage{tabularx}
\usepackage{ltablex} 
\usepackage{longtable}

\usepackage{supertabular}

\usepackage{placeins}
\usepackage{pgffor}

\usepackage{color}
\usepackage{comment}
\usepackage{enumitem}

\usepackage{morefloats} 
\usepackage{chngcntr} 

\usepackage[utf8x]{inputenc}
\usepackage{amsmath}
\usepackage{url}
\usepackage{footnote}

\usepackage[autostyle]{csquotes}  
\usepackage{booktabs} 

\usepackage{hyperref}

\begin{document}
%
\title{Collective Attention towards Scientists and Research Topics}

\author{Claudia Wagner}
\affiliation{%
  \institution{GESIS - Leibniz Institute for the Social Sciences and U. of Koblenz-Landau}
  \city{Cologne/Koblenz}
\country{Germany} }
\email{claudia.wagner@gesis.org}

\author{Olga Zagovora}
\affiliation{%
  \institution{GESIS - Leibniz Institute for the Social Sciences}
  \city{Cologne}
\country{Germany} }
\email{olga.zagovora@gesis.org}

\author{Tatiana Sennikova}
\affiliation{%
  \institution{U. of Koblenz-Landau}
  \city{Koblenz}
\country{Germany} }
\email{tsennikova@uni-koblenz.de}

\author{Fariba Karimi}
\affiliation{%
  \institution{GESIS - Leibniz Institute for the Social Sciences}
  \city{Cologne}
\country{Germany} }
\email{fariba.karimi@gesis.org}

\begin{abstract}
Emergent patterns of collective attention towards scientists and their research may function as a proxy for scientific impact which traditionally is assessed via committees that award prizes to scientists. Therefore it is crucial to understand the relationships between scientific impact and online demand and supply for information about scientists and their work. 
In this paper, we compare the temporal pattern of information supply (article creations) and information demand (article views) on Wikipedia for two groups of scientists: scientists who received one of the most prestigious awards in their field and influential scientists from the same field who did not receive an award.

Our research highlights that awards function as external shocks which increase supply and demand for information about scientists, but hardly affect information supply and demand for their research topics.
Further, we find interesting differences in the temporal ordering of information supply between the two groups: (i) award-winners have a higher probability that interest in them precedes interest in their work; (ii) for award winners interest in articles about them and their work is temporally more clustered than for non-awarded scientists. 
\end{abstract}


\maketitle

\begin{figure*}[t!]
\centering
\begin{minipage}[b]{0.49\textwidth}
    \includegraphics[width=\linewidth, trim=11 10 9 10, clip=true]{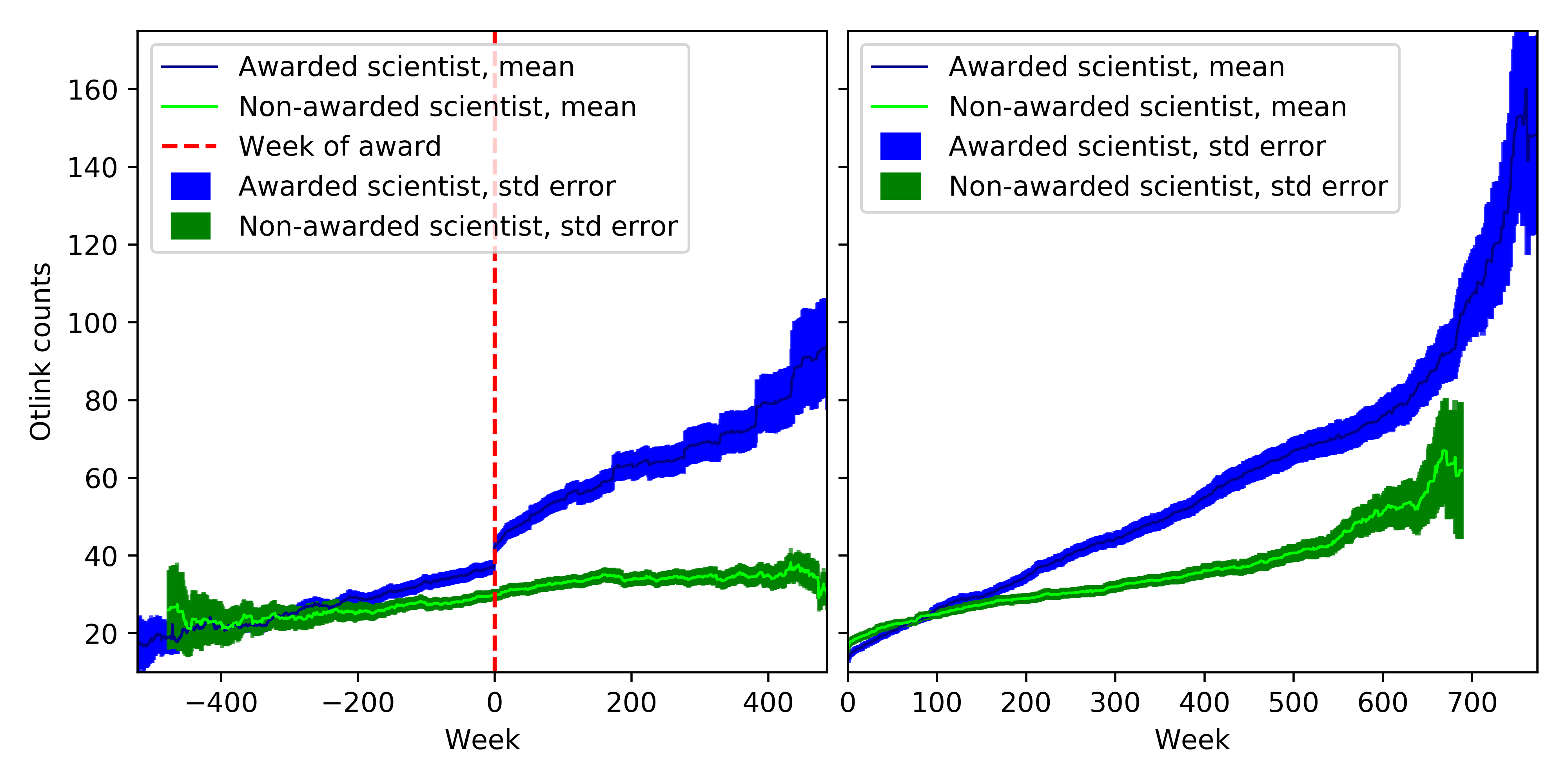}
    \subcaption{Outgoing link counts} \label{1:a}
\end{minipage}
\begin{minipage}[b]{0.49\textwidth}
    \includegraphics[width=\linewidth, trim=11 10 11 10, clip=true]{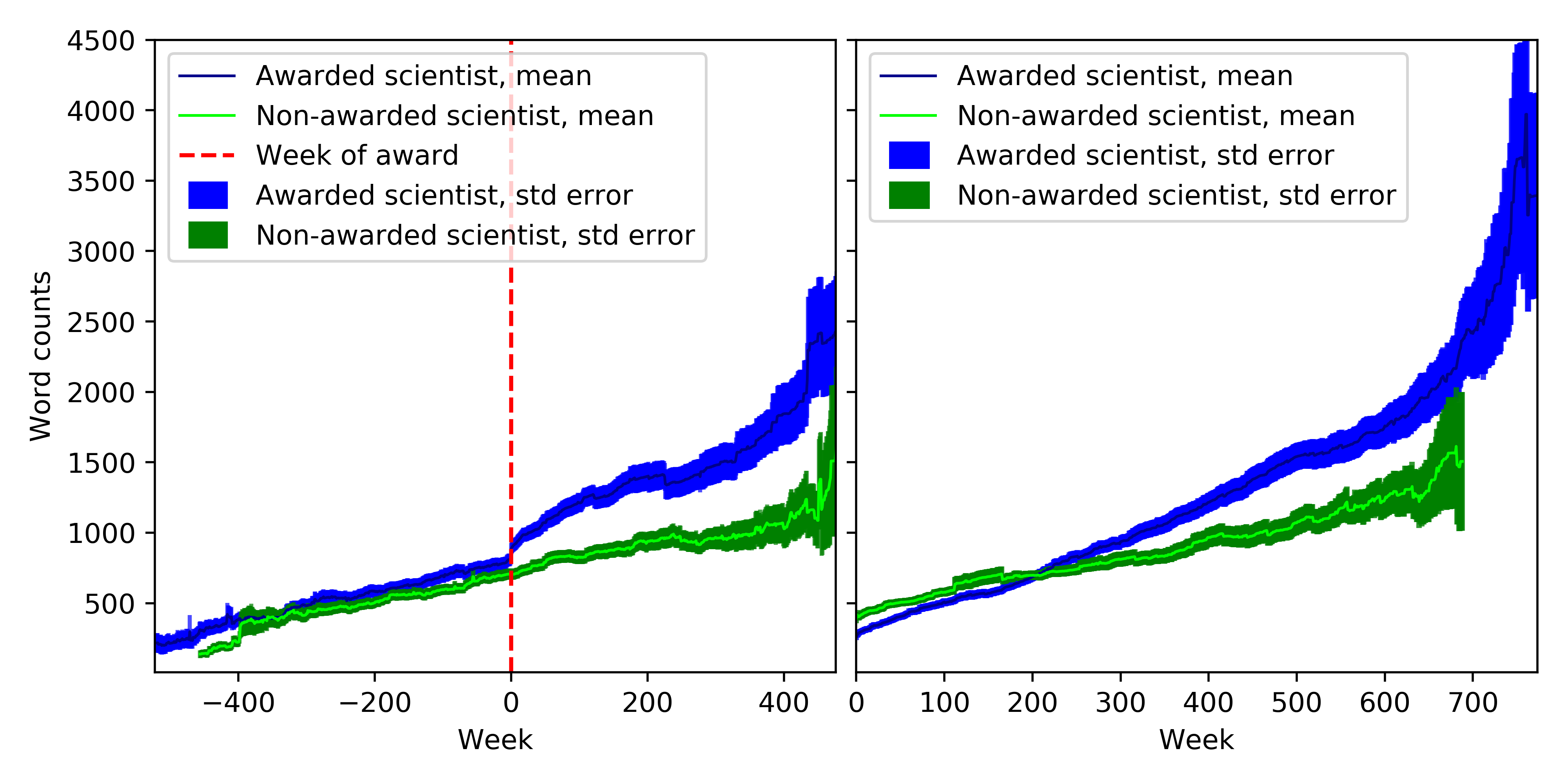}
    \subcaption{Word counts} \label{1:b}
\end{minipage}
\caption{\textbf{Cumulative growth of articles on Wikipedia:} article length is measured via outgoing links in subfigure (\ref{1:a}) and via word counts in subfigure (\ref{1:b}). The zero point refers either to the week when the scientist was awarded (dashed red line) or the week when the article about the scientist was created (in plots without dashed red line). For the non-awarded scientists we picked a random week out of the range during which awards happened (i.e. between 2008-03-27 and 2015-10-12) as placebo points. One can see a discontinuity in the growth of outlinks and words that is related with the award. 
}
\label{fig1}
\end{figure*}

\begin{figure}[t!]
\centering
\begin{minipage}[b]{0.49\columnwidth}
    \includegraphics[width=\linewidth, trim=7 7 5 7, clip=true]{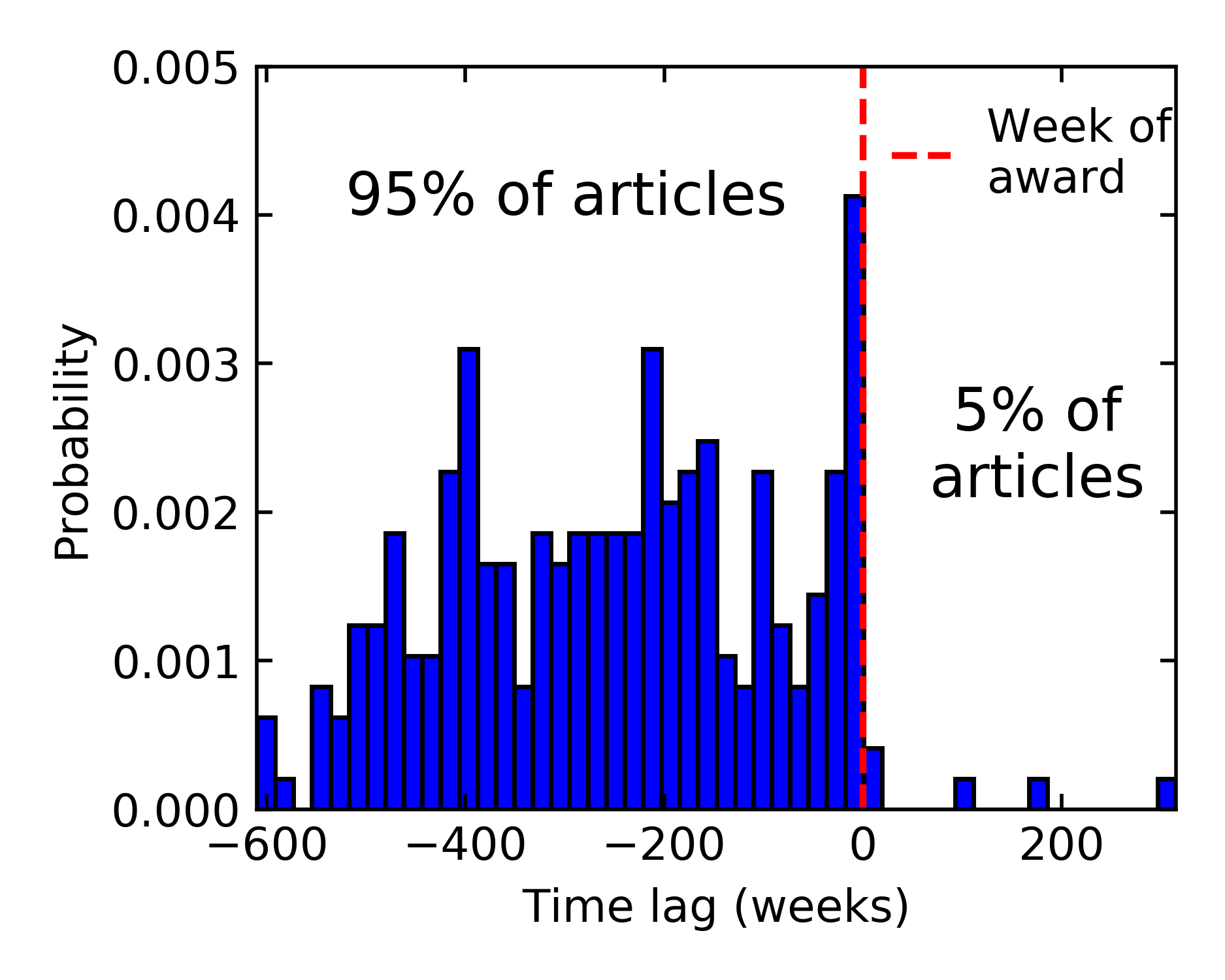}
    \subcaption{Scientists}
    \label{2:a}
\end{minipage}
\begin{minipage}[b]{0.49\columnwidth}
    \includegraphics[width=\linewidth, trim=7 7 5 7, clip=true]{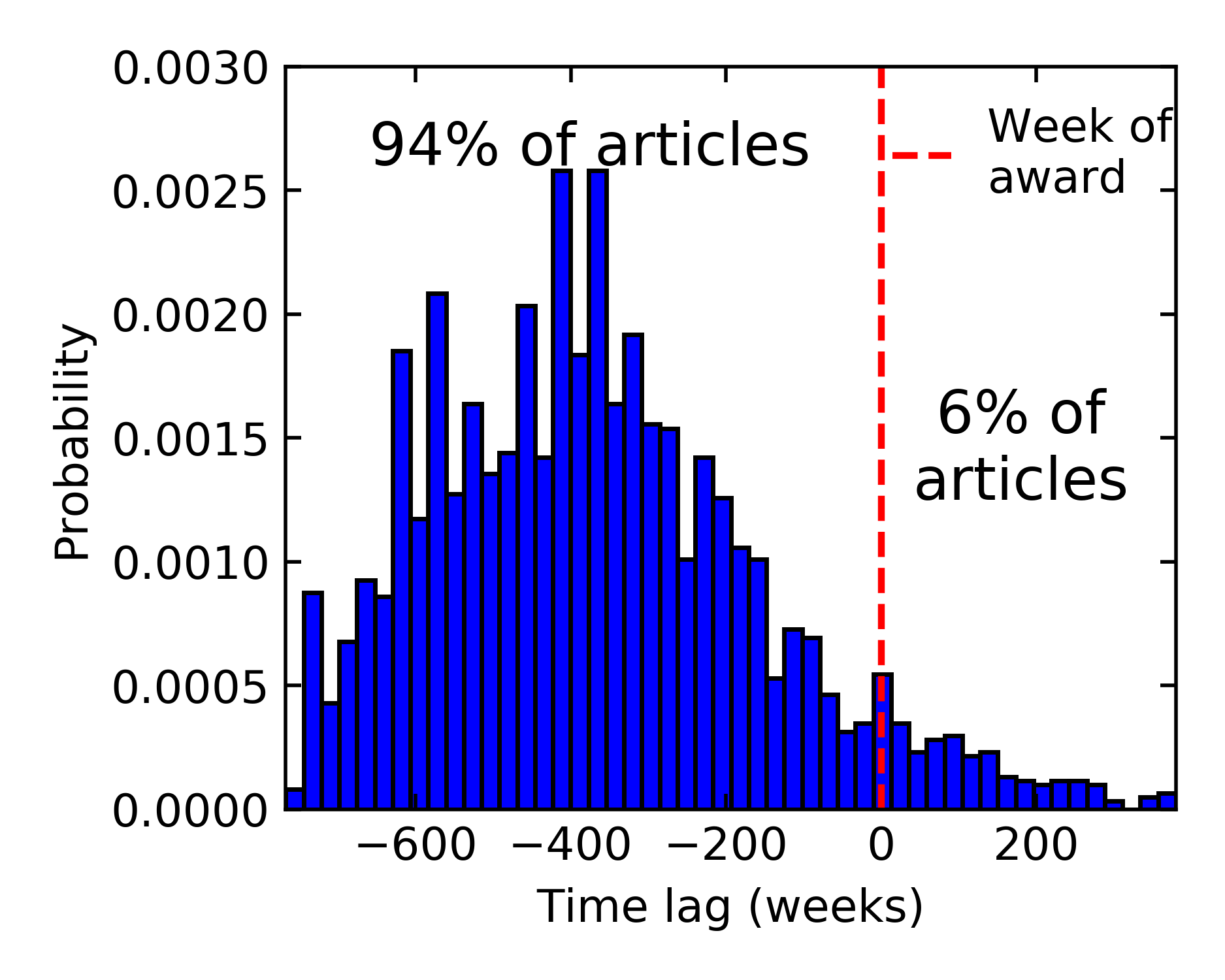}
    \subcaption{Research topics}
    \label{2:b}
\end{minipage}

\caption{\textbf{Time lag between the award and the creation date for articles} about awarded scientist in subfigure (\ref{2:a}) and for their research topics in subfigure (\ref{2:b}). The zero point on the x-axis refers to the week of the award. Most articles about awarded scientists and their research topics have been created before they received the award. 
}
\label{fig2}
\end{figure}

\section{Introduction}

The temporal dynamics of online information supply and demand~\cite{Ciampaglia2015production} for research topics and scientists may reveal information about their impact. 
For example, if a scientist innovates a new research topic, the interest of the general public in the topic will be most likely connected with the interest in the scientists (or the other way around). 
Therefore the interest will be temporally clustered. If interest in the scientist increases, the topic will probably also gain interest or vice versa. 
Conversely, if a scientist's works on a research topic had attracted attention from the general public long before anyone was interested in the scientist, then the interest in the research topic was not driven by the scientist since the temporal order is a necessary (but not a sufficient) condition for causality. 

In this work, we compare the temporal patterns of information supply (article creations) and information demand (article views) on Wikipedia for two groups of scientists: scientists who received one of the most prestigious awards in their field and influential scientists that work in the same field but did not receive an award.


Our research highlights that awards function as external shocks which increase information supply and demand about scientists, but hardly affect the demand and supply for information about research topics.
Though 95\% of articles about scientists have been created before they received an award, information supply is impacted by awards since we find a discontinuity in the growth patterns during the time of the award. After the award, articles about award-winners start to grow much faster than those of non-awarded scientists, while the growth pattern is identical for both groups before that day.

Further, we find interesting differences in the temporal ordering of information supply about scientists and their research topics within the two groups: for award winners information supply about scientists and their research topics is temporally more clustered than for non-award winners. 
That means for award-winners articles about their research topics are created around the same time as the article about the scientist, while for non-award winners larger time-lags are observed.
Award-winners also have a higher probability that interest in them precedes interest in their work,
while for non-awarded scientists, 90\% (for award-winners only 73\%) of the articles about their research topics have been created before the article about the scientist was created.
It is not surprising that for both groups the majority of topics were described on Wikipedia before the articles about the scientist was created, since ``normal science'' is cumulative \cite{Kuhn1970}. That means most scientists work on research topics that have attracted attention in the past.
But for award-winners we find more exceptions; 27\% of their research topics become of interest to the general public after the scientists attracted attention on Wikipedia. One potential explanation is that award winners may innovate new topics or work on relatively new topics.
Examples of Wikipedia articles about research topics that were created after the article of the scientists, provide anecdotal evidence for this explanation: Bayesian Networks after Judea Pearl, Public key cryptography after Whitfield Diffie and Martin Hellman, and Tablet PC after Charles P. Thacker.



To our best knowledge, this is the first study that investigates the impact of scientific awards on the production and consumption of information on Wikipedia.  
We hope that this work is relevant for the Altmetrics community since it sheds light on the temporal dynamics of supply and demand for information about scientists and their research topics online.





\begin{figure*}[h!]
\begin{minipage}[b]{0.33\textwidth}
    \includegraphics[width=\linewidth, trim=10 10 10 10, clip=true]{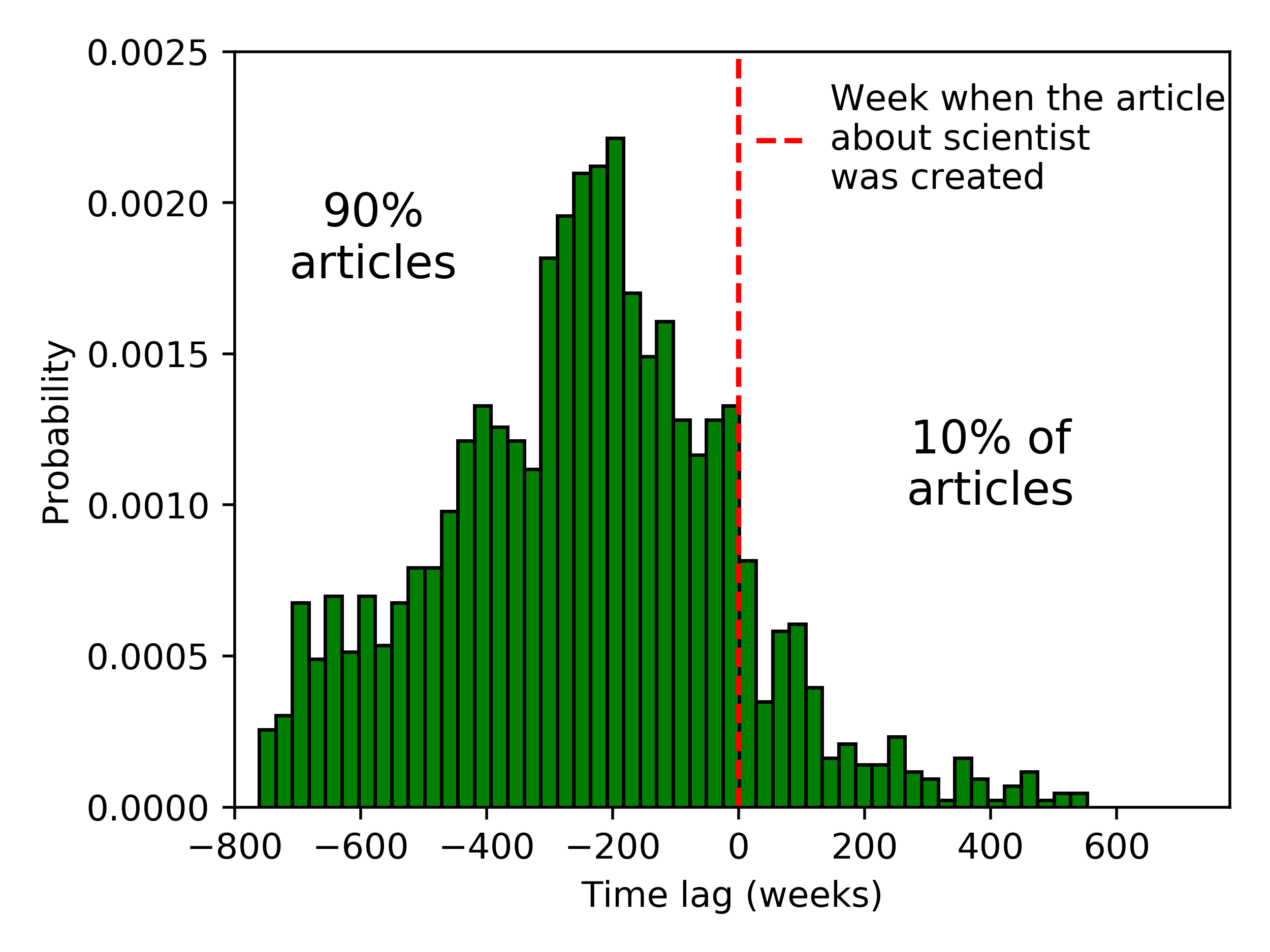}
    \subcaption{Non-awarded group }
    \label{3:a}
\end{minipage}
\begin{minipage}[b]{0.33\textwidth}
    \includegraphics[width=\linewidth, trim=10 10 10 10, clip=true]{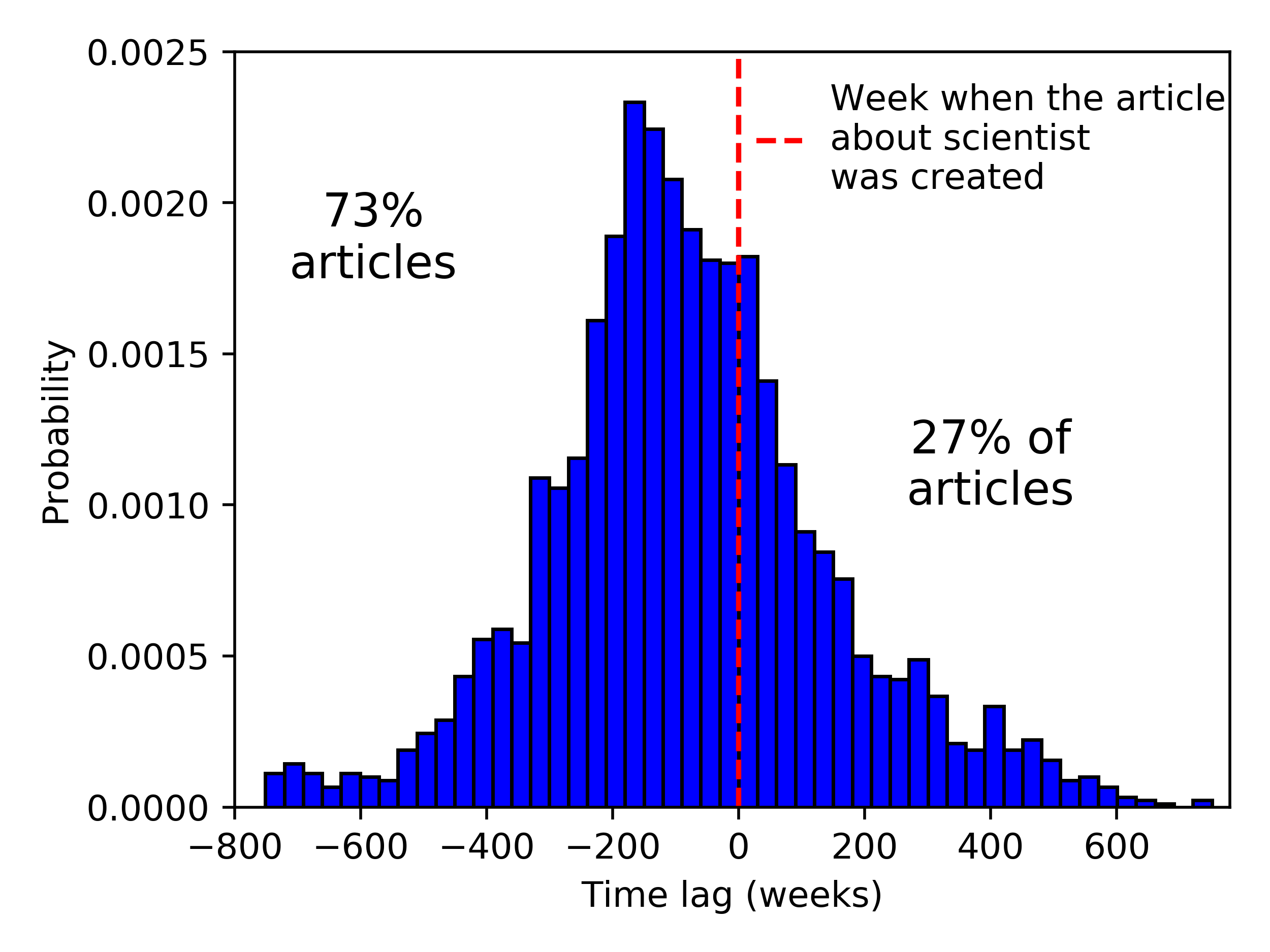}
    \subcaption{Awarded group}
    \label{3:b}
\end{minipage}
\begin{minipage}[b]{0.33\textwidth}
    \includegraphics[width=\linewidth, trim=6 2 24 18, clip=true]{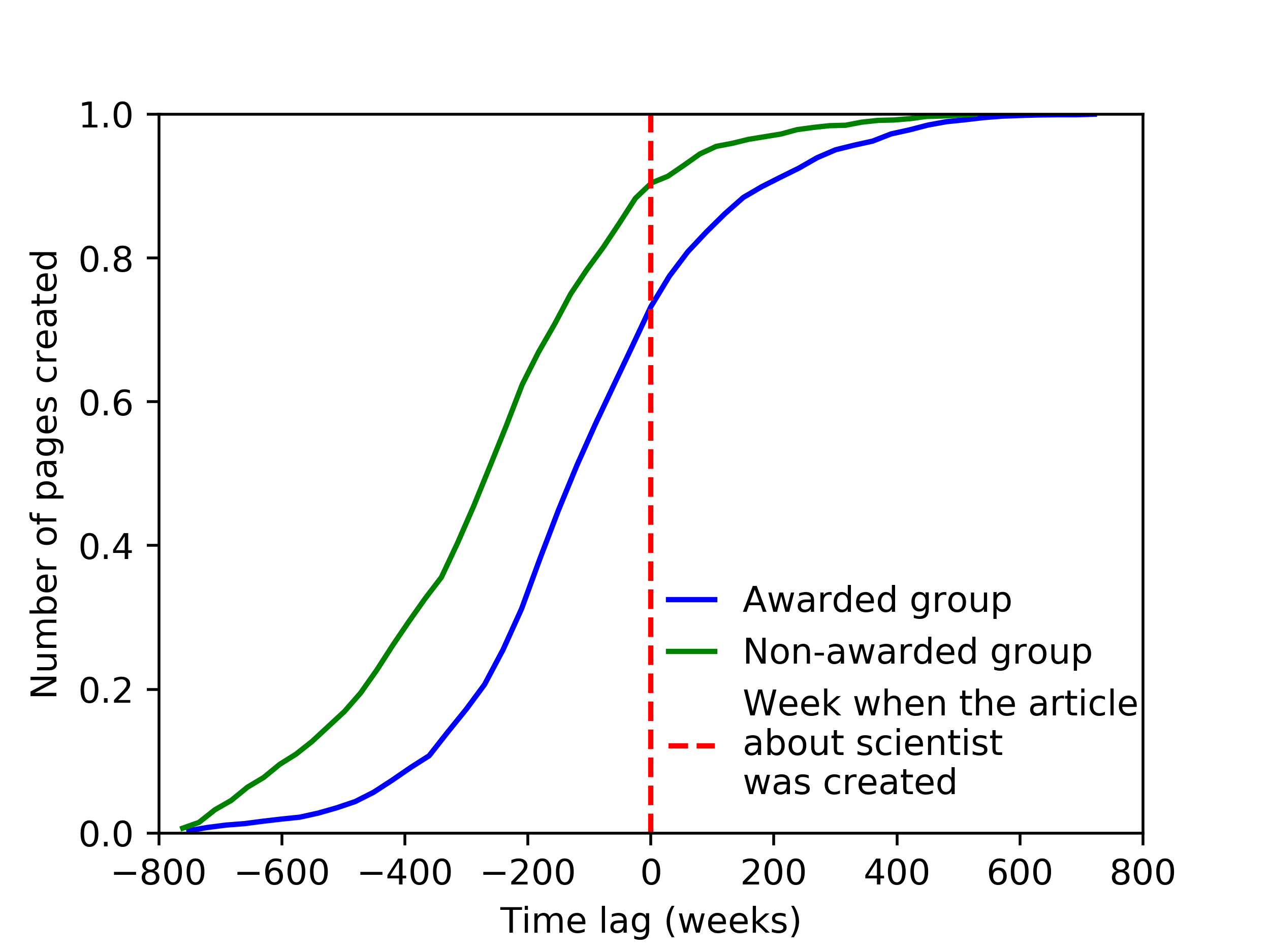}
    \subcaption{CDFs of time-lags}
    \label{3:c}
\end{minipage}

\caption{\textbf{Time-lags between creation dates of Wikipedia articles about scientists and their research topics.} The zero point on the x-axis corresponds to the week when the article about the scientist was created. 
Subfigure (\ref{3:a}) shows that the time-lag is negative for 90\% of all articles about research topics of non-awarded scientists.
This suggests that most articles about research topics have been created before the articles about the non-awarded scientists. 
For awarded-scientists we see a similar pattern in (\ref{3:b}), but the fraction of articles about research topics that are created after the article about the scientist was created is higher (25\%) for awarded scientists than for non-awarded scientists (10\%).
Subfigure (\ref{3:c}) shows that articles about research topics of non-awarded scientists are created earlier than those of awarded scientists relative to the creation date of articles about scientists.
}
\label{fig3}
\end{figure*}

\section{Data and Methods}

We use Wikipedia article creation dates, edits and views as a proxy for online attention.
All data were collected in August 2016. Our code, stopwords list, and datasets are available online\footnote{\url{https://github.com/tsennikova/scientists-analysis} (accessed Apr.~11,~2018) and \linebreak \url{https://github.com/gesiscss/scientists-analysis-wikipedia}  (accessed Apr.~11,~2018) }.
Both datasets of awarded and non-awarded scientist contain the same number of academics from different fields: 57 Physicists, 18 Mathematicians, 18 Computer Scientists, 50 Chemists, 58 Medicine and Physiology researchers, 9 Biologists, and 54 Economists. All together, there are 262 unique researchers in each dataset.

\textbf{Awarded scientists: }\label{seed_scientists}
This dataset focuses on scientists from the aforementioned fields whose work was honoured through some of the most prestigious academic prizes. We consider the awards between 2008 and 2015, since the Wikipedia page view statistics are not available for earlier years. 
Within this time frame, we compile a list of winners of the following prizes and awards: Nobel Prize (77 winners), Abel Prize (10), Fields Medal (8), Turing Award (10), IEEE Medal of Honor (8), and International Prize for Biology (9). We also include 163 Thomson Reuters Citation Laureates\footnote{\url{http://stateofinnovation.thomsonreuters.com/hall-of-citation-laureates} (accessed Jul.~28,~2016)} (23 of whom also received the Nobel Prize), which are selected for outstanding contributions based on the citation impact of their published research. 
We manually map these winners to the corresponding Wikipedia articles in the English edition, and record their scientific field, gender, award year, and the date when the Wikipedia article was created. 
The final sample consists of 262 awarded scientists and is available online\footnote{\url{https://github.com/tsennikova/scientists-analysis/blob/master/data/seed/seed_creation_date.json} (accessed Apr.~11,~2018)}.

\textbf{Non-awarded Scientists: }\label{sec:baseline-scientists}
For a fair comparison, we select a sample of influential, highly cited scientists who worked at the same time, in the same scientific fields as the award winners, using the Thomson Reuters database of Highly Cited Scientists\cite{reuters}.
We use all records between 2001 and 2015 and 
remove scientists who have received an award. Finally,  we draw a random stratified sample of 262 academics with the same distribution across scientific fields as in the awarded dataset. We also map these researchers to articles in the English Wikipedia and add information about their scientific field and the date when the Wikipedia article was created (available online\footnote{\url{https://github.com/tsennikova/scientists-analysis/blob/master/data/baseline/baseline_creation_date.json} (accessed Apr.~11,~2018)}).

\textbf{Scientific Topics: }\label{sec:related_topics}
For all researcher in our sample we analyze their Wikipedia article and construct a list of scientific topics related to the scientist. For that, we extract all outgoing links from the articles about scientists in the English Wikipedia.
Each of these articles has a category section (found at the bottom of the page) which displays a subject area of the article, and helps readers to navigate through related concepts. 
We use this concept list and a manually created set of stop words (available online\footnote{\url{https://github.com/tsennikova/scientists-analysis/blob/master/data/neighbors/stop_list.txt} (accessed Apr.~11,~2018)}) to remove articles that are related with a scientists but are not related to research areas (e.g. locations, institutions).
We evaluate our filtering approach by comparing the algorithmic assessment with a manual assessment for 10 randomly selected articles about scientists and all outgoing links from these articles.
The evaluation results show that our filtering method is very effective: the overall accuracy is 0.96, precision is 0.93, and recall is 0.9.
Overall, we construct a list of 1,911 topics\footnote{\url{https://github.com/tsennikova/scientists-analysis/blob/master/data/neighbors/seed_neighbors_list_clean_en.json} (accessed Apr.~11,~2018)} that are related to awarded scientists and 1,070 topics\footnote{\url{https://github.com/tsennikova/scientists-analysis/blob/master/data/neighbors/baseline_neighbors_list_clean_en.json} (accessed Apr.~11,~2018)} that are related to non-awarded scientists.


\textbf{Wikipedia page views: }
We collect daily page views of all articles about scientists and their research topics.
We use page view statistics from the project Wiki Trends\cite{wikitrends}, which itself is based on the Wikimedia data dumps\cite{wikidumps}. Wiki Trends data provides aggregated number of daily visits to Wikipedia articles and all redirects to them, collected from the English edition.
In order to eliminate influences of daily and seasonal fluctuations of article views, we normalize the data as follows:
\begin{equation}
\bar V_{i,d} =\frac {V_{i,d} * max(M)}{M_d}.
\end{equation}
where $V_{i,d}$ refers to the number of visits to an article $i$ on day $d$, $M_{d}$ is the number of Wikipedia Main Page views for the same day, and $max(M)$ is the maximum number of Wikpedia Main Page views.
We collect page views that happened between 01.01.2008 and 01.05.2016.




\section{Results}

\smallskip
\textbf{Information supply: }
First we explore how collective attention on  Wikipedia is affected by external events such as awards, looking at the temporal order of article creations and edits on Wikipedia. 
Figure \ref{fig1} shows that articles for awarded and non-awarded scientists grow similarly before the award. But the award creates a discontinuity since articles of awarded scientists start to grow faster than those of their non-awarded colleagues; more hyperlinks and more words are added. 
This suggests that the award triggers additional information supply, though most articles about awarded scientists and their research topics are created before they receive the award (see Figure \ref{fig2}).

\begin{figure*}[t!]
\begin{minipage}[b]{0.49\textwidth}
    \includegraphics[width=\linewidth, trim=10 10 11 11, clip=true]{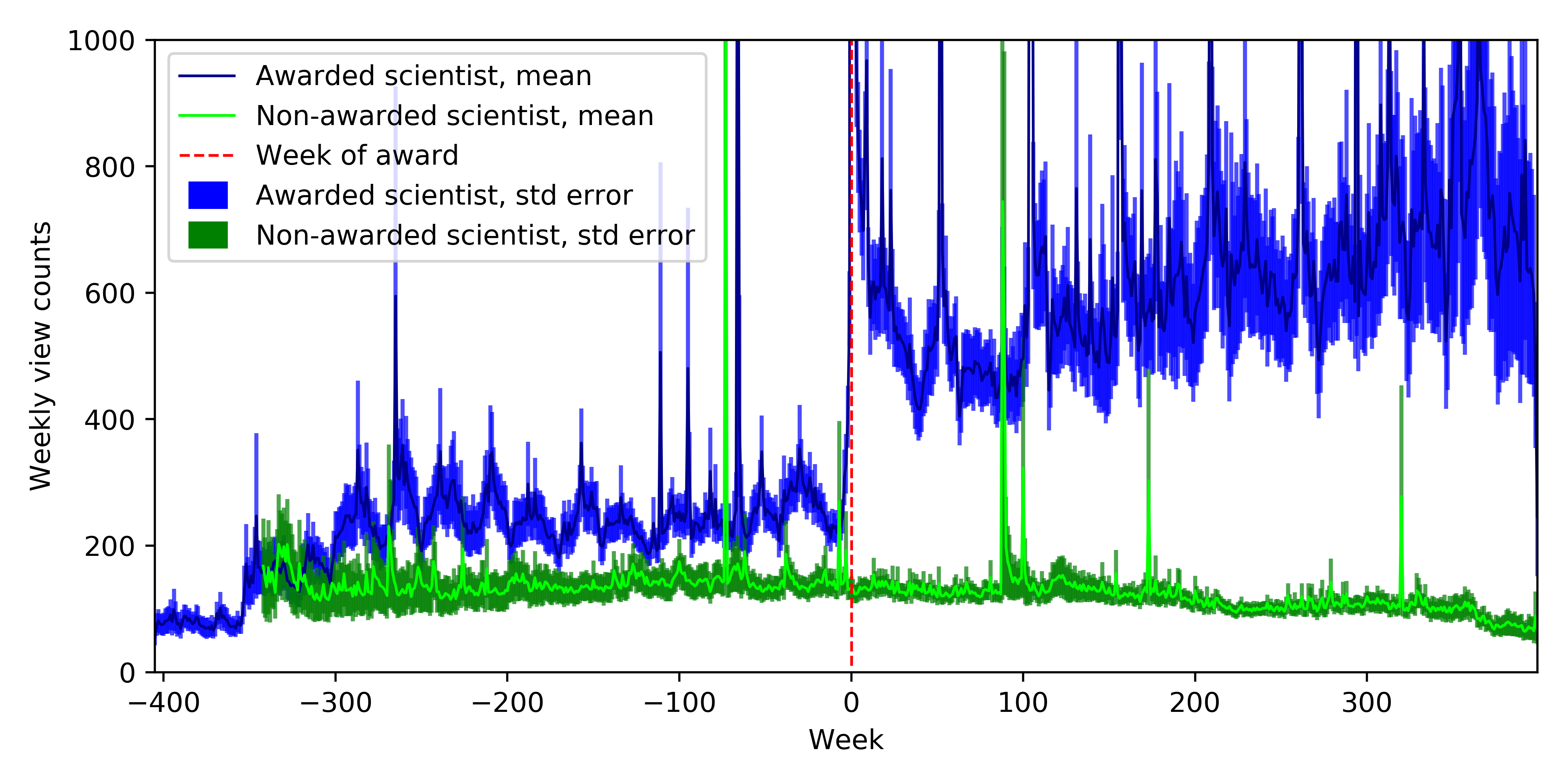}
    \subcaption{View counts for scientists}
    \label{4:a}
\end{minipage}
\begin{minipage}[b]{0.49\textwidth}
    \includegraphics[width=\linewidth, trim=10 10 11 11, clip=true]{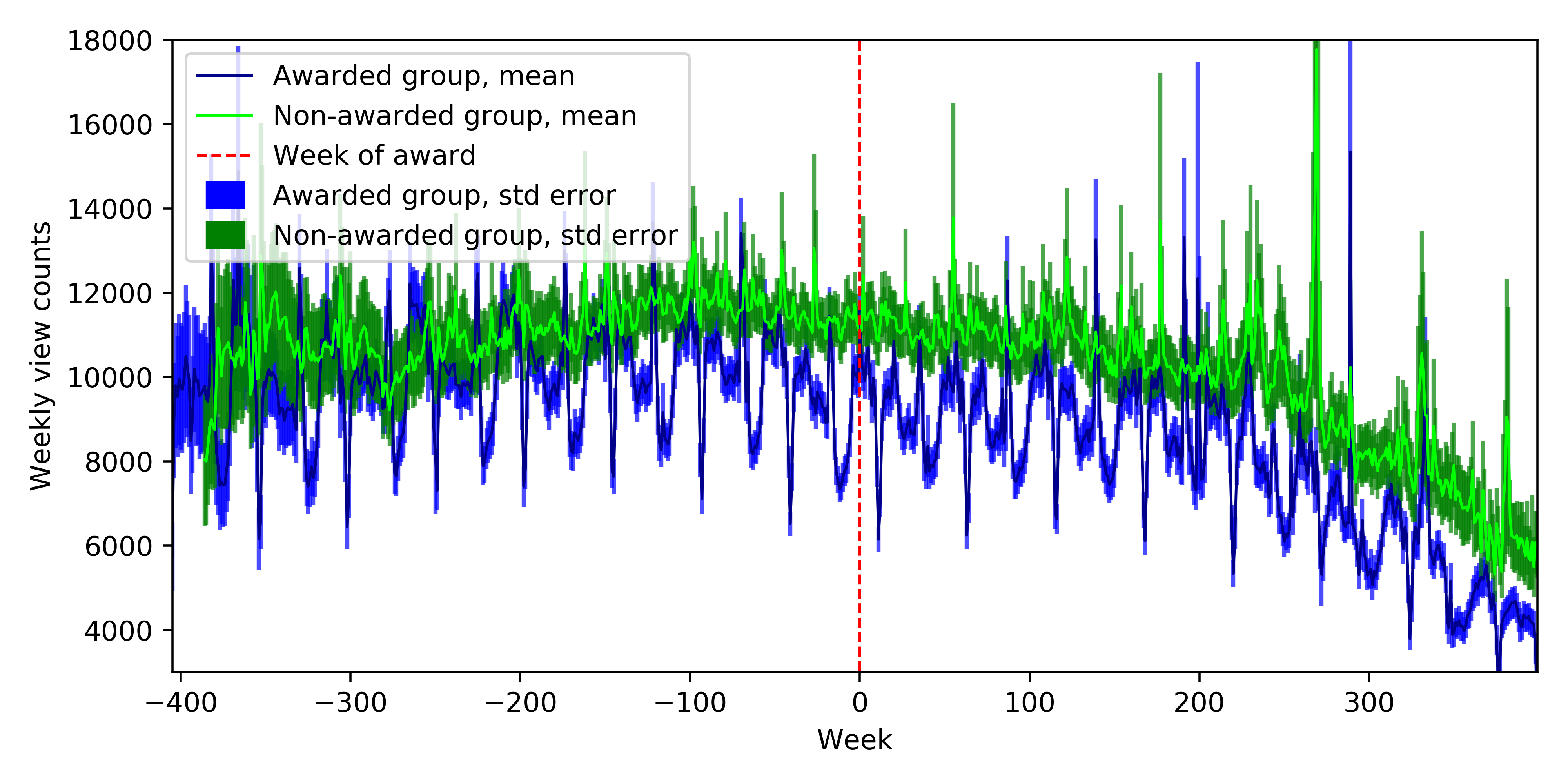}
    \subcaption{View counts for research topics}
    \label{4:b}
\end{minipage}
\caption{\textbf{Weekly view counts for articles  about scientists and research  topics.} The zero point refers to the week when the scientists was awarded (dashed red line). For the non-awarded scientists we picked a random week out of the range during which awards happened (i.e. between 2008-03-27 and 2015-10-12) as placebo points. One can see that the information demand on scientists is clearly impacted by the award, however the interest in research topics associated with the scientists seems to be unaffected.}
\label{fig4}
\end{figure*}

\smallskip
But what came first: the interest in the scientist or the interest in her research? To address this question, we examine the time lag between the article creation about scientists and the scientific topics associated with them.
We compare the differences in weeks between the creation dates. The time lag is positive if the topic article was created after the article about the scientist, and it is negative otherwise. 
Figure \ref{fig3} shows the probability density function of time lags for both groups of scientists.
The zero point on the x-axis refers to the week when the article about the scientist was created. 
One can see that for both groups most articles about research topics that are associated with a scientist are created before the article about the scientist is created. 
That means, information supply for research topics usually precedes information supply for scientists.
This is not surprising since ``normal science'' is cumulative \cite{Kuhn1970} and most scientists work on research topics that have attracted attention in the past.
But award-winners have a higher probability that interest in them precedes interest in their work. For award winners 27\% of articles about research topics related to the scientist have been created after the article about the scientist was created, while for non-award winners only 10\% of the articles have been created after the article about the scientist was created.
One potential explanation for this difference is that award winners are more likely to innovate new topics or work on relatively new topics and therefore articles about these topics have not yet been created.

Figure \ref{fig3} also shows that the dispersion of the time-lag distribution for award winners is lower which means that the temporal distances between the creation dates of articles about award-winners and their research topics vary less than for non-award winners. This suggests that interest in scientists and their research topics is more interrelated for award-winners than for non-award winners.


\smallskip
\textbf{Information demand: }
So far we have seen that awards impact the production of new information on Wikipedia. Articles about scientists grow faster after they receive an award. However, it remains unclear how the consumption of information is affected by the award. 
Is the demand for information about scientists and their research topics increasing after they win an award? And how long-lasting is this effect?

Figure \ref{fig4} shows that the information demand for scientists is impacted by the award, since we see a clear discontinuity in the view counts for articles about scientists who won an award.
For non-awarded scientists we pick a random day out of the range during which awards happened (i.e. between 2008-03-27 and 2015-10-12) as placebo points to compute a baseline. 
The baseline indicates how much change we would expect to see by chance. The discontinuity which we see in Figure \ref{fig4} clearly goes beyond what we would expect by chance. Also the increased information demand seems to remain rather stable over time. Even 300 weeks after the award, we see that the information demand for awarded scientists is on average higher than those for non-awarded scientists.
Interestingly, we see that the information demand for articles about research topics associated with the award-winners seems to be unaffected by the award (cf. Figure \ref{4:b}).

\section{Related Work}
Quantifying and predicting scientific success is a topic of high interest for the academic community \cite{sinatra,penner,fortunato_science_2018,bar-ilan_beyond_2012}. While it is clear that online attention to scientists does not always coincide with their academic rigor \cite{samoilenko_distorted_2014}, more research is needed to understand the reasons of such inconsistencies, and the meaning behind them. 

Work on collective attention has mainly focused on the consumption of information \cite{WU2007,hodas_how_2012,Lilian2012srep,widen_chapter_2012} and has shown information consumption correlates with real-world events, such as the spread of influenza \cite{Ginsberg2009}, box office returns \cite{Mestyan2013} and scientific performance \cite{Shen12325}. 

Only recently researchers started exploring the interplay between information production and consumption. In \cite{Ciampaglia2015production}
the authors show that the production of new information on Wikipedia is associated with significant shifts of collective attention measured via article views.  
That means, in many cases, demand for information precedes its supply. However, unexpected events may lead to almost instantaneously article creations which are followed by a short period of high information demand.
A scientific award can be an expected or an unexpected event. Therefore, it is unclear if new articles about scientists will be created on Wikipedia directly after the award, even if no changes in information demand are observed before the award. Our work shows that in most cases articles about scientist and research topics precede the award. 
However, we see that awards boost the demand for information about scientists and that the increased demand lasts over the next few years. 
\section{Discussion}

How does an award impact information supply and consumption online?
If awards would be totally unexpected and hit scientists randomly, they would lead to almost instantaneous article creations which would be followed by a short increase in information demand~\cite{Ciampaglia2015production}.
Our work shows a different pattern and suggests that awards are probably not so unexpected and have long term effects. 95\% of the articles about scientists are created before they receive an award, but information supply is impacted by awards since articles about award-winners grow faster than those of non-awarded scientists. Also information supply is impacted by the award since we find a discontinuity in the view counts in the week when the prize was awarded.
Interestingly the increased information demand is rather stable over time. Even five years after the award, we see that the information demand for awarded scientists is on average higher than the demand for non-awarded scientists.

The discontinuity which we see during the week when the prizes are awarded suggests that awards may have a causal effect on information production and consumption. 
However, to establish a hard causal link future research is necessary since other factors that correlate with awards may exist and confound our analysis.

We also find interesting differences between the two groups of scientists when looking at the information production side. For award-winners articles about them and their research topics are temporally more clustered than for non-award winners. Award-winners also have a higher probability that interest in them precedes interest in their work. One potential explanation is that award winners are more connected with their research topic since they may innovate new topics or work on relatively new topics. 
Examples of topic articles that were created after the article of the scientist provide anecdotal evidence for this explanation: Bayesian Networks after Judea Pearl, Public key cryptography after Whitfield Diffie and Martin Hellman, and Tablet PC after Charles P. Thacker.
However, further research is necessary to explore the different types of relationships between scientists and their research topics that will lead to the creation of a hyperlink on Wikipedia.



\section{Conclusions}

The goal of this work was to understand the impact of awards on the production and consumption  of information about scientists and their research topics on Wikipedia.

Our work shows that (i) scientists who win a prestigious prize attract more attention afterwards (i.e., information supply and demand increases but only for articles about scientists); (ii) information supply for award winners and their research topics is temporally more clustered than for non-award winners; and (iii) information supply for award winners is more likely to precede information supply for their research topics compared to non-award winner.


For future work it would be interesting to extend this group level analysis with an individual level analysis and further investigate the different types of relationships between scientists and their research topics that may lead to a hyperlink on Wikipedia.
We also collected gender information about scientists and plan to compare the information supply and demand for female and male award-winners and influential scientists that did not receive an award in future work.

\section{Acknowledgments}
This work is part of the DFG-funded research project \textit{*metrics} (project number: 314727790). Further information on the project can be found at \url{https://metrics-project.net}.




\bibliographystyle{ACM-Reference-Format}
\bibliography{bibliography}

\end{document}